\documentclass[12pt]{article}
\usepackage{graphicx,amsmath,cite,subfig,authblk}

\usepackage[a4paper, total={6in, 8in}]{geometry}

\begin{document}
\title{Integrals of motion in 3-d Bohmian  Trajectories}      

\author{A.C. Tzemos\footnote{thanasistzemos@gmail.com} \, and   G. Contopoulos\footnote{gcontop@academyofathens.gr}}

\affil{Research Center for Astronomy and Applied Mathematics of the Academy of Athens - Soranou Efessiou 4, GR-11527 Athens, Greece}

\date{}
\maketitle

\vspace{7pt}

\begin{abstract}
Chaos in Bohmian Quantum Mechanics is an open field of research. In general, most of the 3-d Bohmian trajectories are free to wander around the 3-d space. However there are cases where the evolution of the trajectories is dictated by exact or approximate integrals of motion. A first case corresponds to partial integrability, where the trajectories (ordered and chaotic) evolve on certain integral surfaces. A second case corresponds to ordered trajectories. In this paper we extend our previous work in 3-d Bohmian Chaos by using both forms of integrability and discuss their physical implications.
\end{abstract}

\section{Introduction}
\label{intro}

In Classical Mechanics chaos describes the sensitivity of the orbits of a nonlinear dynamical system to the initial conditions.
In the standard approach of  Quantum Mechanics (SQM) the linearity of Schr\"{o}\-din\-ger's equation and the absence of the notion of trajectory makes the definition of quantum chaos ambiguous both conceptually and quantitatively \cite{gutzwiller2013chaos,wimberger2014nonlinear}.

However, there is a great debate in the last 90 years about alternative interpretations of QM. One of the oldest alternative interpretations is Bohmian Quantum Mechanics (BQM) \cite{debroglie1927a,debroglie1927b,Bohm,BohmII,benseny}. In BQM the wavefunction $\psi$, which is  the solution of Schr\"{o}\-din\-ger's equation, acts as a pilot wave \cite{debroglie1927a, debroglie1927b} that guides the quantum particles on well defined deterministic trajectories. BQM is closely related to the Madelung's hydrodynamical approach of the probability fluid \cite{madelung1927quantentheorie, trahan2005quantum}. In short, BQM is a quantum theory with trajectories, where  the theory of classical dynamical systems fully applies.

General applications of the  Bohmian trajectories have been presented in the study of several classes of physical systems, such as  molecular dynamics \cite{C0CP02175J}, Bose-Einstein condensates \cite{fetter2001vortices}, Josephson junctions \cite{bruder1999phase},  light-matter interaction \cite{lai2009quantum}, nanoelectronics \cite{albareda2013time}, decoherence \cite{sanz2007quantum}, interferometry \cite{sanz2002particle} etc. In recent years a number of books and review papers  on BQM and its applications have been published \cite{holland1995quantum, trahan2005quantum, durr2009bohmian, sanz2012trajectory, sanz2013trajectory,benseny}.

Chaos in BQM has attracted a lot of interest in the last two decades, since it is important both for theory and applications. A detailed bibliography  is given in previous papers of ours \cite{efthymiopoulos2006chaos, efthymiopoulos2017chaos}. Some basic examples of its utility are:
\begin{enumerate}
\item{The study of the  relation of classicaly integrable/chaotic systems and their quantum counterparts: The nonlinear but deterministic character of Bohmian equations of motion, makes BQM ideal for the study of    order and chaos in quantum systems, with the same mathematical techniques used in classical mechanics. We note here that, as stated by Benseny et al. \cite{benseny}, unlike classical mechanics, in BQM it is  difficult to detect integrals of motion. Furthermore in quantum chaos emerges from the complexity of the quantum potential (namely from the wavefunction) rather than from the external potentials.}
\item{The dynamics of  many-body systems whose complexity increases exponentially with the number of particles. BQM and its close relation to quantum hydrodynamics (QHD) \cite{trahan2005quantum} has been used in order to effectively simulate large systems by means of low dimensional systems. In QHD one observes often  the existence of vortical flows around the nodes of the wavefunctions. Quantum vortices are strongly related to chaotic behavior \cite{frisk1997properties, falsaperla2003motion, wisniacki2007vortex, Efth2009} and have been detected in processes like atom-surface scattering  \cite{sanz2004role}, wavepacket dynamics \cite{chou2009hydrodynamic}, electron scattering through crystalls \cite{efthymiopoulos2012wavepacket} etc. }
\item{Quantum relaxation \cite{VALENTINI19915, VALENTINI19911, valentini2005dynamical, towler2012time} is the dynamical approach to  Born's rule $\rho=|\psi|^2$ when particles are started with an initial distribution $\rho_{initial}\neq|\psi_{initial}|^2$. It has been emphasized that  chaos is a necessary condition in order  for quantum relaxation to effectively proceed  \cite{contopoulos2012order}. However, we note that the emergence of Born's rule in BQM is a controversial subject (see e.g. \cite{durr1992quantum}) and different approaches have been presented in the literature  \cite{durr2009bohmian}. }
\end{enumerate}

Thus, the BQM trajectory approach opens a new field related to quantum phenomena, where all the methods and tools of the theory of classical dynamics  can be applied. In particular, our goal in the present paper is to show how the Bohmian trajectories allow to unambiguously transfer the notion of integrability (complete or partial)  from the classical to the quantum domain.

The mechanism responsible for the emergence of chaos in Bohmian trajectories is an open problem of BMQ.  Discussions of order and chaos in 2-d systems have been made by several authors up to now (\!\!\cite{frisk1997properties,wisniacki2005motion,efthymiopoulos2006chaos,efthymiopoulos2007nodal,Efth2009,contopoulos2008ordered,contopoulos2012order}  and references therein).
Bohmian 2-d chaos occurs when a trajectory passes close to a 'nodal point-X-point complex', a characteristic geometrical structure of the Bohmian flow, composed of a moving nodal point (defined as the solution of $\Psi=0$ at any given time $t$) and a second stationary (saddle) point of the flow,  called the X-point. In a previous paper \cite{contopoulos2008ordered} we have shown that ordered orbits satisfy, in general, an approximate integral of motion in the form of a formal series. Such orbits avoid close encounters with the nodal point-X-point complexes of the quantum flow.

In the present paper we extend our study to 3-d systems, where little work has been made up to now \cite{falsaperla2003motion, wisniacki2005motion, wisniacki2007vortex, Tzemos2016,contopoulos2017partial}. In particular we extend the notions of order and chaos in a 3-d system of quantum oscillators in the presence/absence of partial integrability. A system is called partially integrable when there exists one (only) exact integral of motion that confines Bohmian trajectories on certain integral surfaces, as shown in \cite{Tzemos2016,contopoulos2017partial}. Partial integrability simplifies significantly the technical difficulties of the computations. On the other hand it implies restrictions for the shape and, consequently, the degree of chaos in the Bohmian trajectories. In the present paper we discuss the role of the integrals of motion, exact or approximate, on the orbits of Bohmian particles.

In section 2 we give the Bohmian equations of motion and the form of the wavefunction used. In Section 3 we present three characteristic examples of partial integrable cases, discuss the trajectories of the nodal points and  find the X-points on the integral surfaces.
In section 4 we study the orbits in these three cases and in a completely nonintegrable case. We find both ordered and chaotic orbits. Most ordered orbits satisfy formal integrals of motion of the ``third integral'' type \cite{Contopoulos200210}, as in the 2-d case. In section 5 we summarize our conclusions. Finally, in the Appendix we discuss some details about the computation of the trajectories of the nodal points.

\section{Equations of motion}
The Bohmian equations of motion \cite{Bohm, BohmII} in 3-d systems are
\begin{align}
m_i\frac{dx_i}{dt}=\Im\Big(\frac{\nabla_i \Psi}{\Psi}\Big)=\frac{1}{G}\Big(\frac{\partial \Psi_I}{\partial x_i}\Psi_R-\frac{\partial \Psi_R}{\partial x_i}\Psi_I\Big),\quad i=1,2,3\label{bohmeqs}
\end{align}
with $G=\Psi_R^2+\Psi_I^2$, where $\Psi$ is a solution of Schr\"{o}dinger's equation $H\Psi=E\Psi$.
In the case of 3-d harmonic oscillators it is convenient to work with a wavefunction $\Psi$
of the form
\begin{align}\label{form}
\Psi(\vec{x},t)=&a\Psi_{p_1,p_2,p_3}(\vec{x},t)+b\Psi_{r_1,r_2,r_3}(\vec{x},t)+c\Psi_{s_1,s_2,s_3}(\vec{x},t),
\end{align}
where $
|a|^2+|b|^2+|c|^2=1,$ $
\Psi_{n_1,n_2,n_3}(\vec{x},t)=\Psi_{n_1,n_2,n_3}(\vec{x})e^{-iE_it/\hbar}$
and $\Psi_{n_1,n_2,n_3}(\vec{x})$ are eigenstates of the 3-d harmonic oscillator of the form
\begin{align}\label{eigenstate}
\Psi_{n_1,n_2,n_3}(\vec{x})=\prod_{k=1}^3\frac{\Big(\frac{m_k\omega_k}{\hbar\pi}\Big)^{\frac{1}{4}}\exp\Big(\frac{-m_k\omega_kx_k^2}{2\hbar}\Big)}{\sqrt{2^{n_k}n_k!}}H_{n_k}\Big(\sqrt{\frac{m_k\omega_k}\hbar}x_k\Big).
\end{align}
$n_1,n_2,n_3$ stand for their quantum numbers, $\omega_1,\omega_2,\omega_3$ for their frequencies and $E_1,E_2,E_3$ for their energies. From now on we set $m_i=\hbar=1$ and write $x_1,x_2,x_3$ as $x,y,z$. For a given set $(n_1,n_2,n_3)$ we have $E=\sum_{i=1}^3(n_i+\frac{1}{2})\omega_i$.

It has been shown \cite{efthymiopoulos2007nodal,Efth2009,Tzemos2016} that chaos is generated when an orbit approaches an unstable point X in the neighbourhood of a nodal point N where the wavefunction $\Psi$ vanishes. A superposition of the form (\ref{form}) with three basis functions presents particular interest, since in this model we can find all three cases of integrable, partially integrable and non-integrable Bohmian trajectories. In fact if we have a sum of more than three basis functions, the system is generally nonintegrable, while if we have a sum of only two basis functions the system is integrable \cite{contopoulos2017partial}.
In the case of three basis functions we have found \cite{contopoulos2017partial} that there are partially integrable cases, where we have an integral of motion of the form $f(x,y,z,t)=C$. The various combinations of quantum numbers that produce partial integrability are given in Table \ref{tab:1}.

\begin{table}[h]
\centering
\caption{Various conditions for the combinations of quantum numbers that produce partial integrable Bohmian trajectories.}\vspace{1cm}
\label{tab:1}       
\begin{tabular}{llll}
\hline\noalign{\smallskip}
Case & first triplet & second triplet & third triplet  \\
\noalign{\smallskip}\hline\noalign{\smallskip}
1& $r_1=p_1$ & $s_2=r_2$ & $s_3=p_3$ \\
\\
2& $r_1=p_1$ & $s_2=p_2$ & $s_3=r_3$ \\
\\
3 &$s_1=r_1$ & $r_2=p_2$ & $s_3=p_3$ \\
\\
4&$s_1=r_1$ & $s_2=p_2$ & $r_3=p_3$ \\
\\
5& $s_1=p_1$ & $r_2=p_2$ & $s_3=r_3$ \\
\\
6& $s_1=p_1$ & $s_2=r_2$ & $r_3=p_3$ \\
\noalign{\smallskip}\hline
\end{tabular}
\end{table}

We will consider now 3 characteristic partially integrable systems. One that has spherical integral surfaces, one with closed integral surfaces (non-spherical) and one with open integral surfaces extending to infinity \cite{Tzemos2016,contopoulos2017partial}.

\section{Partially Integrable Cases}
\label{sec:1}

\subsection{Spherical integral surfaces}
\begin{figure}[hb]
\centering
\includegraphics[scale=0.32]{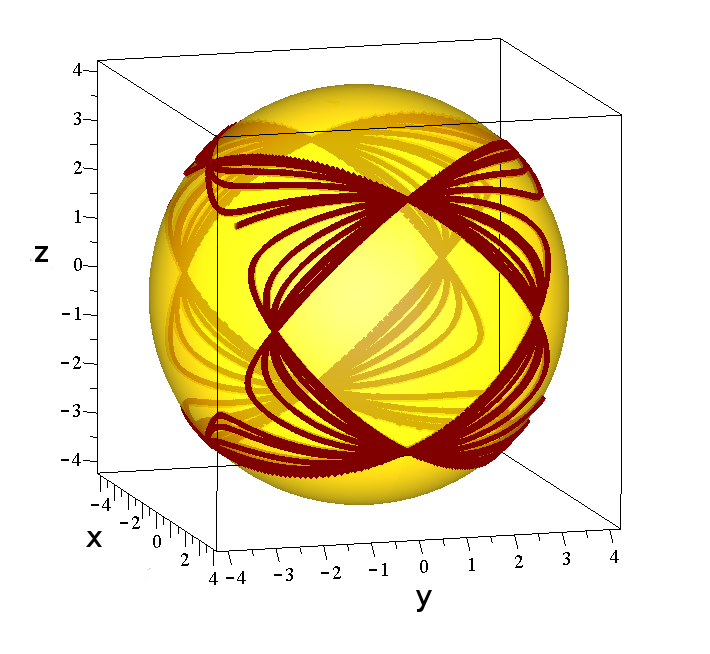}
\caption{The trajectory of a nodal point on a certain spherical surface $(R=4.234)$ in 3-d space. We observe the special points of convergence on the plane  $z=0$ and on the upper and lower hemisphere for $x=0$ and for $y=0$.}
\label{np1}
\end{figure}

The most simple case is
\begin{align}
\Psi(\vec{x},t)=a\Psi_{100}(\vec{x},t)+b\Psi_{010}(\vec{x},t)+c\Psi_{001}(\vec{x},t)
\end{align}
In this case the integral surfaces are spheres
\begin{align}
x^2+y^2+z^2=R^2=C
\end{align}
The orbit of the nodal point on a spherical surface is found from the equations \cite{Tzemos2016}
\begin{align}\label{nodp}
x_{nod}(t;R)=\frac{S\sin(\omega_{32}t)}{a\sqrt{\omega_1}},\quad
y_{nod}(t;R)=\frac{S\sin(\omega_{13}t)}{b\sqrt{\omega_2}},\quad
z_{nod}(t;R)=\frac{S\sin(\omega_{21}t)}{c\sqrt{\omega_3}},
\end{align}
where
\begin{align}
S=\frac{R}{\sqrt{\frac{\sin^2(\omega_{32}t)}{{a^2\omega_1}}+\frac{\sin^2(\omega_{13}t)}{{b^2\omega_2}}+\frac{\sin^2(\omega_{21}t)}{c^2\omega_3}}}
\end{align}
with $\omega_{ij}=\omega_i-\omega_j,  (i,j=1,2,3)$, $x_{nod}^2+y_{nod}^2+z_{nod}^2=R^2$. We set  $a=b=c=\frac{1}{\sqrt{3}}$ so that $|a|^2+|b|^2+|c|^2=1$.
In particular if $z_{nod}=0$ we have $\sin(\omega_{21}t)=0$, i.e. $t=\frac{K\pi}{\omega_{21}}$ with integer $K$. Then 
\footnote{$\sin\Big(\frac{\omega_{13}}{\omega_{21}}K\pi\Big)=
\sin\Big(\frac{\omega_{23}-\omega_{21}}{\omega_{21}}K\pi\Big)=
\sin\Big(\frac{-\omega_{32}}{\omega_{21}}K\pi-K\pi\Big)=\pm \sin\Big(\frac{\omega_{32}}{\omega_{21}}K\pi\Big)$}
\begin{align}
\frac{y_{nod}}{x_{nod}}=\sqrt{\frac{\omega_1}{\omega_2}}\frac{\sin\Big(\frac{\omega_{13}}{\omega_{21}}K\pi\Big)}{\sin\Big(\frac{\omega_{32}}{\omega_{21}}K\pi\Big)}=\pm\sqrt{\frac{\omega_1}{\omega_2}}
\end{align}
Therefore for $z_{nod}=0$ there are only 4 directions $\phi=\pm\arctan\Big(\sqrt{\frac{\omega_1}{\omega_2}}\Big)$, i.e 4 points on the plane $z=0$ of the sphere. Similarly there are 4 points on the plane $x=0$ and 4 points on the plane $y=0$ (Fig.~\ref{np1}).

\begin{figure}
\centering
\subfloat{\includegraphics[scale=0.35]{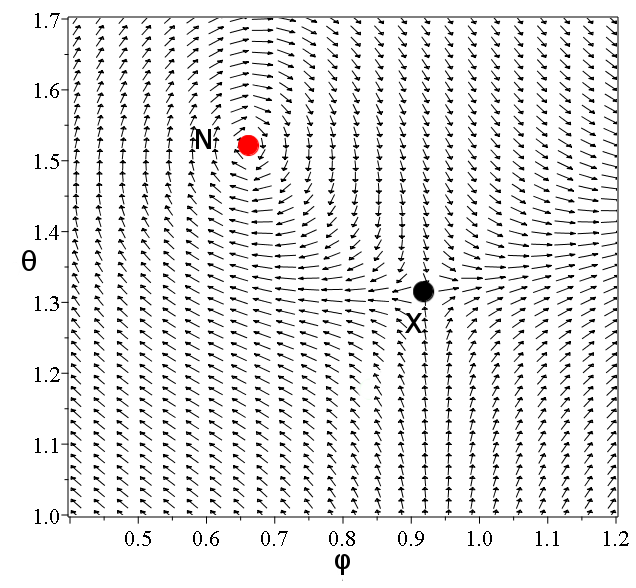}}
\subfloat{\raisebox{2.5cm}{\includegraphics[scale=0.2]{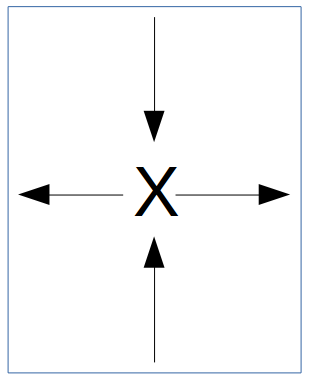}}}
\caption{A stereographic projection  of the Bohmian flow on a sphere. Close to the nodal point (N) the flow is spiral, while close to the X-point there are two approaching opposite directions (close to the stable manifolds of the X-point) and two deviating opposite directions (close to the  unstable manifolds) as we can see in the side figure.}
\label{npk1}
\end{figure}

In past studies \cite{efthymiopoulos2007nodal, Efth2009, contopoulos2012order} we have demonstrated that the features of quantum trajectories can be studied conveniently by computing the quantum flow in a frame of reference co-moving with the nodal point.
Furthermore, close to the nodal point an `adiabatic approximation' holds, namely the motions are much faster than the evolution of the instantaneous structure of the quantum flow.
The key remark is that close to the nodal point there is an unstable point X that generates chaos \cite{Tzemos2016}. In the present paper we define the X-point on the integral surface. This can be seen in a figure depicting the flow on the surface of the sphere in coordinates $\phi, \theta$ (Fig.~\ref{npk1}) in the frame of reference co-moving with the nodal point. Orbits approaching the X-point are deviated and chaos is introduced. The X-point is in general close to the nodal point and forms with it a 'nodal point-X-point complex' \cite{efthymiopoulos2007nodal,Efth2009}. In the following we draw the trajectories of the nodal points as representative of the nodal point-X-point complex.

\begin{figure}[ht]
\centering
\includegraphics[scale=0.6]{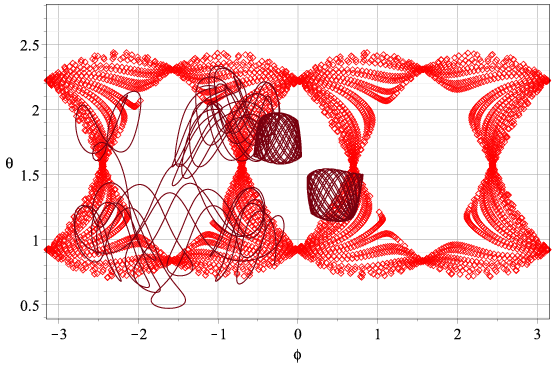}
\caption{The stereographic projection of the trajectory of the nodal point on  the sphere. Here we use spherical coordinates $\phi, \theta$ for a value $R=3$ of the radius of the sphere  for $t\in[1,200]$ and superimposed we give two ordered (box) orbits  (upper orbit: $\phi(1)=0.01,\theta(1)=1.7,$ lower orbit $\phi(1)=0.68,\theta(1)=1.2$) and two chaotic orbits (upper orbit: $\phi(1)=-1,\theta(1)=2,$ lower orbit $\phi(1)=-0.3,\theta(1)=0.8$).}
\label{np2}
\end{figure}

\begin{figure}[h]
\centering
\includegraphics[scale=0.6]{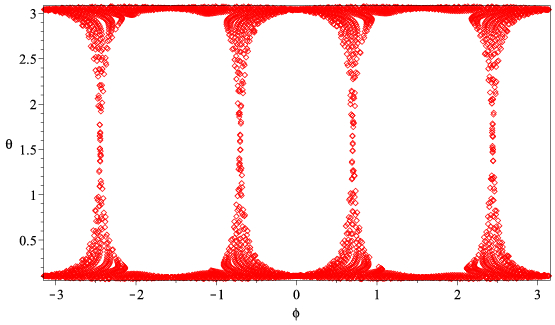}
\caption{The stereographic projection with $c=0.0545$ and $R=3$. As $c\to0$ it results to the isolation of the nodal point on the two poles of the sphere ($\theta=0$ and $\theta=\pi/2$). The concentration of nodal points is then reduced to 4 vertical lines and there is no chaos.}
\label{np3}
\end{figure}
If we draw the trajectory of the nodal point on the spherical coordinates $\phi,\theta$ we get the stereographic projection of Fig.~\ref{np2}. In this figure also we see the regions filled by the trajectory of the nodal points and the empty regions in between. The position of the X-point is always close to the nodal point of the same time \cite{Tzemos2016}.

If now the  coefficient $c$ is reduced the regions filled by the trajectory of the nodal point are reduced too and the empty regions become larger (Fig.~\ref{np3}).
If finally $c\to 0$ the empty regions cover the whole space and the limiting trajectories of the nodal point become 4 vertical lines, while the upper part of Fig.~\ref{np3} becomes a horizontal line representing the pole $\theta=0$ ($\phi$ is undefined). In fact if $\Psi=a\Psi_{p_1p_2p_3}+b\Psi_{r_1r_2r_3}$  the problem is reduced to one dimensional, therefore it is completely integrable and there is no chaos. Moreover, if we have coexistence of two cases among the cases 1-6  of Tab.~\ref{tab:1} the system is again completely integrable.

\subsection{Pear-shaped integral surfaces}

\begin{figure}[hbt]
\centering
\includegraphics[scale=0.5]{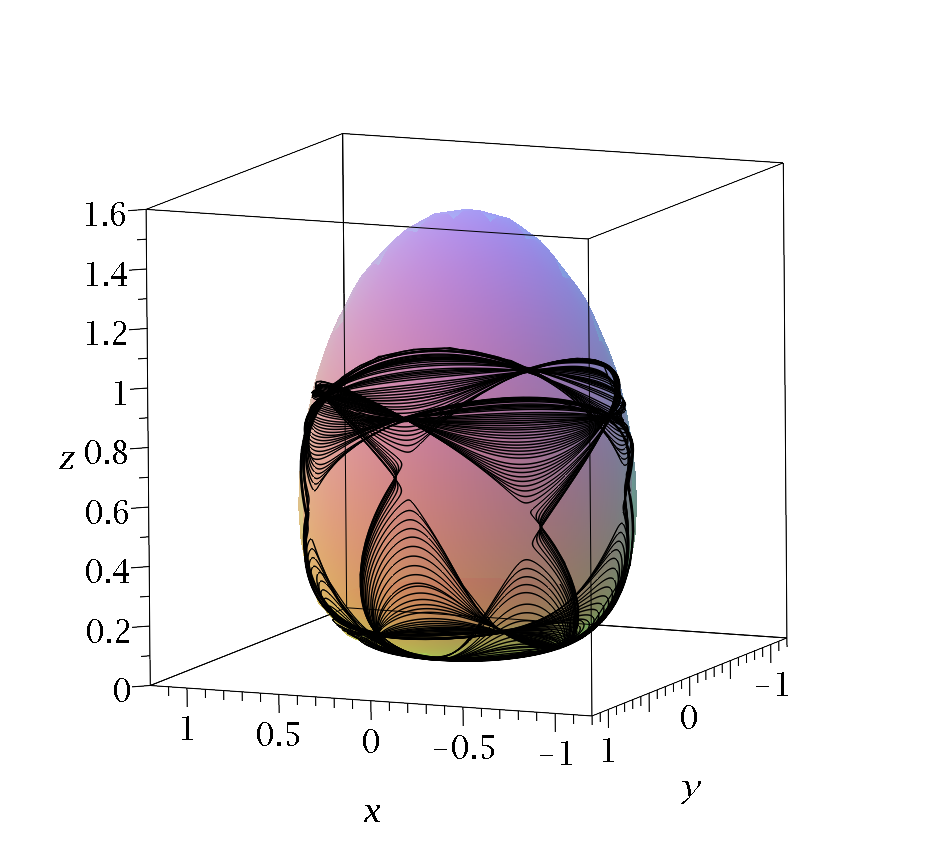}
\caption{The trajectory of a nodal point on a pear shaped integral surface. ($C=1, \omega1=1, \omega_2=\sqrt{2}, \omega_3=\sqrt{3})$}
\label{np4x}
\end{figure}

In this case
\begin{align}
\Psi(\vec{x}, t)=a\Psi_{100}(\vec{x},t)+b\Psi_{010}(\vec{x},t)+c\Psi_{002}(\vec{x},t),
\end{align}
with $a=b=c=1/\sqrt{3}$.
The integral surfaces are
\begin{align}\label{eq_pear}
x^2+y^2+\frac{z^2}{2}-\frac{ln|z|}{2\omega_3}=C
\end{align}
(we take $\omega_3=\sqrt{3}$). The surface consists of two closed surfaces, one with $z>0$ and a symmetric one with respect to the plane $z=0$ (with $z<0$). The two surfaces are completely separated and the orbits cannot move from one surface to the other. The upper part is shown in Fig.~\ref{np4x} and it looks like a pear with a nearly flattened bottom close to $z=0$. This figure is axially symmetric. It contains a nodal trajectory which is similar to the nodal trajectory of the spherical case. In particular we find
\begin{align}
\frac{y_{nod}}{x_{nod}}=\pm\sqrt{\frac{\omega_1}{\omega_2}},\,\, \text{when}\,\,
z_{nod}=\pm\frac{1}{\sqrt{2\omega_3}}
\end{align}
(instead of $z_{nod}=0$  of the spherical case). The trajectory of the nodal point leaves empty domains on the integral surface where we find ordered orbits.

\begin{figure}[h]
\centering
\subfloat{\includegraphics[scale=0.26]{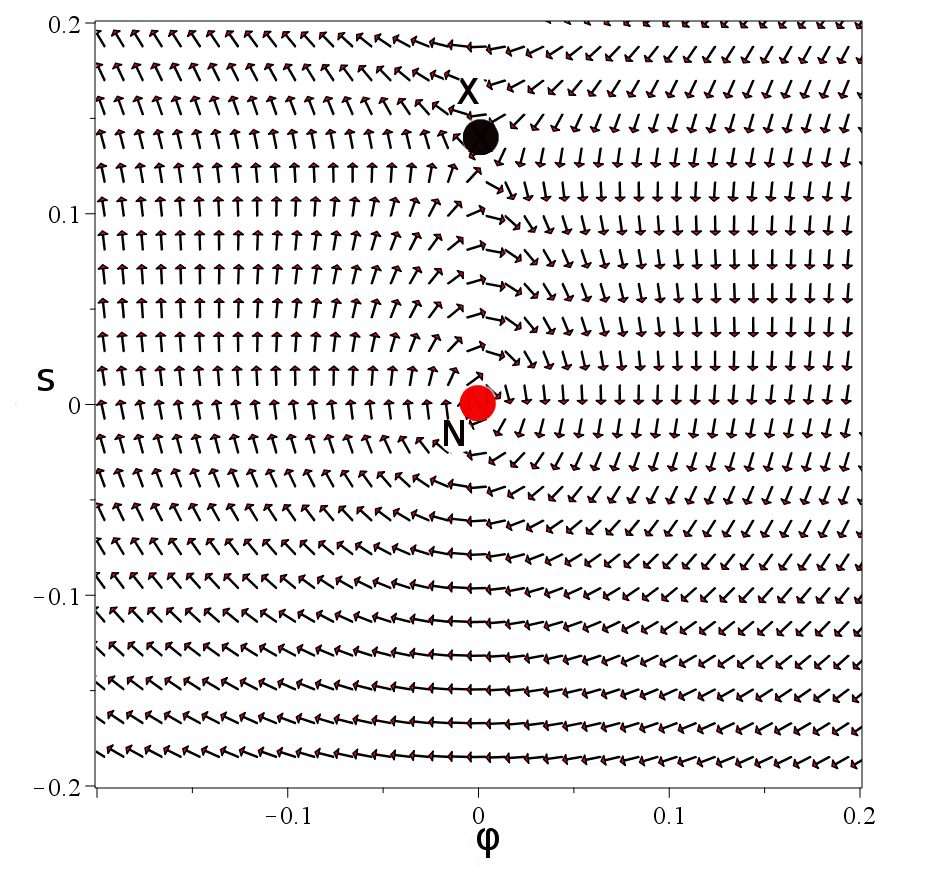}}
\subfloat{\raisebox{2.5cm}{\includegraphics[scale=0.3]{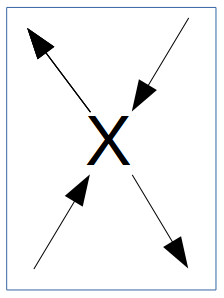}}}
\caption{The nodal point-X-point complex on a pear shaped surface  with respect to the $s,\phi$ coordinates and the directions of the orbits around it. In the side figure we see the directions of the flow at the X-point.}
\label{np6}
\end{figure}

In this case we can find coordinates similar to the coordinates $(\theta, \phi)$ on the sphere of case I. In fact we may write Eq.~\ref{eq_pear} in the form $\rho^2=C-\Phi(z)$
where $\rho^2=x^2+y^2$ and
\begin{align}
\Phi(z)=\frac{z^2}{2}-\frac{\ln|z|}{2\omega_3}
\end{align}
Then we find $ds=\sqrt{d\rho^2+dz^2}=\sqrt{1+\frac{\Phi'(z)^2}{4(C-\Phi(z))}}dz$
and after integration
\begin{align}
s=\int_{z=z_{nod}} ^z\sqrt{1+\frac{\Phi'(z)^2}{4(C-\Phi(z))}}dz
\end{align}
where $z_{nod}$ is the z-coordinate of the nodal point which plays the role of the origin of the new frame of coordinates.
Every point of an orbit has coordinates $(s, \phi=\arctan{(\frac{y}{x})})$ similar to the coordinates $(\theta,\phi)$ of the spherical case.
The directions of the velocities of the various points with respect to the nodal point are given in Fig.~\ref{np6}. We see that the X-point is an unstable point that generates chaotic deviations of the orbits that approach it.

The function $\Phi$ exists for $|z|\geq z_0$, where $z_0$ is the root of the equation $\Psi'=z-\frac{1}{2\omega_3z}=0\Rightarrow z_0=\frac{1}{\sqrt{2\omega_3}}$.
If $\omega_3=\sqrt{3}$ we have $z_0=0.5373$. The corresponding minimum value of C is $C=0.3237$. For any given $C>C_0$ we have two corresponding values of $z$, namely the roots of the equation $\Phi=C$. The maximum value $z_{max}$ is at the top of Fig.~\ref{np4x} for $x=y=0$ and the minimum value $z_{min}$ is at its bottom ($z_{min}>0$ is close to $z=0$).

\subsection{Open integral surfaces}
Such a case is given by the wavefunction
\begin{align}
\Psi(t)=a\Psi_{000}(t)+b\Psi_{110}(t)+c\Psi_{102}(t)
\end{align}
In this case the integral surfaces are
\begin{align}
-x^2+y^2+\frac{z^2}{2}-\frac{\ln|z|}{2\omega_3}=C
\label{open}
\end{align}
and the surfaces extend to infinity.
The intersections of the surface by a given plane z are two hyperbolas $y^2-x^2=C-\Phi(z)$.

\begin{figure}[hbt]
\centering
\includegraphics[scale=0.28]{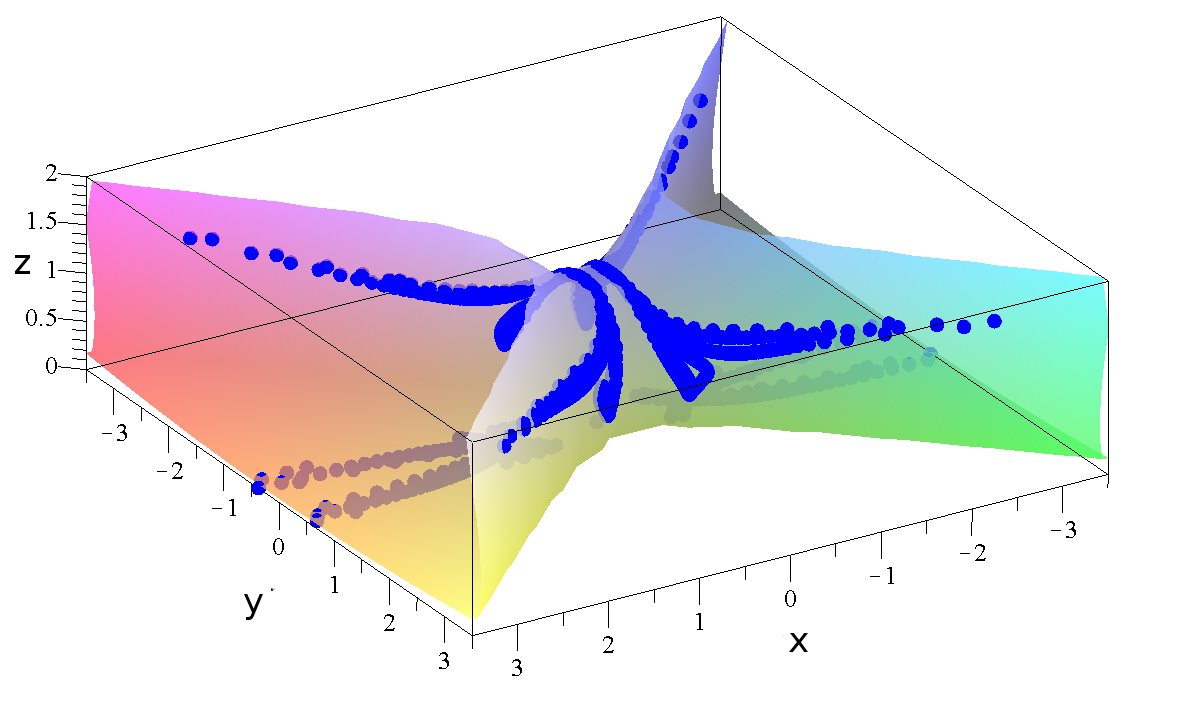}
\caption{The nodal point trajectory on an open integral surface for $z>0$, with $C=1$. The nodal point goes to infinity at certain times along the `arms' of the hyperbolic space and returns from the opposite direction.}
\label{npt}
\end{figure}

If $z_{min}<z<z_{max}$ we have $C-\Phi(z)>0$ therefore $y^2-x^2>0$, i.e. the hyperbolas are above and below the x-axis. On the other hand if $z>z_{max}$ we have $C-\Phi(z)<0$ and $y^2-x^2<0$, i.e. the hyperbolas are on the left and on the right of the y-axis. Similarly if $0<z<z_{min}$ we have again $y^2-x^2<0$ and the hyperbolas are on the left and on the right.
Thus the surface C contains hyperbolas above and below the x-axis that are joined along the diagonals $y^2=x^2$ for $z=z_{max}$ and $z=z_{min}$. The general form of the surface (\ref{open}) for $C>C_0$ is shown in Fig.~\ref{npt}. In this figure we have drawn also the trajectory of the nodal point. This trajectory extends to infinity along particular directions. Similar figures appear for $z<0$. On the other hand if $C<C_0$ there are no $z_{max}$ and $z_{min}$ and we have always $C-\Phi(z)<0$ therefore we have always $y^2-x^2<0$ i.e. a surface consisting only of hyperbolas on the left and on the right.

\section{Orbits}
The orbits of particles are found by solving numerically the equations of motion. We use the adaptive Runge-Kutta-Fehlberg 4-5th order numerical scheme for all of our numerical integrations, with absolute error tolerance equal to $10^{-7}$ and relative error tolerance equal to $10^{-6}$. Thus our results are not disturbed qualitatively by the errors, even in the more difficult case  of chaotic trajectories.
\subsection{Partially Integrable Cases}
\subsubsection{Spherical Surfaces}

In the case of the spherical integral surface  we find by use of  Eqs. (\ref{bohmeqs})
\begin{align}
&\frac{dx}{dt}=\frac {-\sin \left( \omega_{12}\,t \right) y\sqrt{\omega_1\omega_2}ab-\sin \left( \omega
_{13}\,t \right) z\sqrt{\omega_1\omega_3}ac}{G}\label{bohmeqs1}\\&
\frac{dy}{dt}=\frac {-\sin \left( \omega_{23}\,t \right) z\sqrt{\omega_2\omega_3}bc+\sin \left( \omega
_{12}\,t \right) x\sqrt{\omega_1\omega_2}ab}{G
}\label{bohmeqs2}\\&
\frac{dz}{dt}=\frac {\sin \left( \omega_{23}\,t \right) y\sqrt{\omega_2\omega_3}bc+\sin \left( \omega_
{13}\,t \right) x\sqrt{\omega_1\omega_3}ac}{G}\label{bohmeqs3}
\end{align}
\begin{align}G=&\nonumber 2\,\cos \left( \omega_{12}\,t \right) xy\sqrt{\omega_1\omega_2}ab+2\,\cos \left( \omega_{13}\,t \right) xz\sqrt{\omega_1\omega_3}ac+2\,\cos \left(
\omega_{23}\,t \right) yz\sqrt{\omega_2\omega_3}bc\\&+{x}^{2}q_1+{y}^{2}q_2+{z}^{2}q_3,\end{align} with
$q_1=a^2\omega_1, q_2=b^2\omega_2, q_3=c^2\omega_3$.
Moreover
$x\dot{x}+y\dot{y}+z\dot{z}=0$. Thus
\begin{align}\label{sfairaol}
x^2+y^2+z^2=R^2.
\end{align}
 We can apply regular perturbation theory by taking $b,c$ sufficiently small and search for approximative solutions as series in $b$ and $c$ in the case of ordered orbits. The solution is in the form of series in powers of $b$ and $c$: $x=x_0+x_1(t)+x_2(t)+\dots$ and similarly for $y$ and $z$. These series can be written in the form $x-x_1(t)-x_2(t)-\dots=x_0$
and similar expressions for $y_0$ and $z_0$, giving formal integrals of motion of the form of the `third integral'\cite{Contopoulos200210}.
Up to first order in $b,c$ the solution is:
\begin{align}
&x=x_0+\frac {[\cos \left(
\omega_{12}\,t \right)-1 ]\sqrt{\omega_1\omega_2}ab\,y_{0}\,\omega_{13}+[\cos \left( \omega_{
13}\,t \right)-1] \sqrt{\omega_1\omega_3}ac\,z_{0}\,\omega_{12}}{\omega_{12}\,\omega_{13}\,q_{1}\,{x_{0}}^{2
}}\label{first}\\&
y=y_0+\frac {[
 1-\cos \left( \omega_{12}\,t\right)]\sqrt{\omega_1\omega_2}ab}{q_{1}\,x_{0}\,\omega_{12}}\label{second}\\&
z=z_0+\frac {[1-\cos \left( \omega_{13}\,t
 \right)] \sqrt{\omega_1\omega_3}ac}{q_{1}\,x_{0}\,\omega_{13}}\label{third}
\end{align}

We note that the solutions (\ref{first}-\ref{third}) have trigonometric terms with only two frequencies $\omega_{12}, \omega_{13}$. Following the arguments of the papers \cite{efthymiopoulos2007nodal,contopoulos2008ordered} we find that the same is true for higher order terms. Furthermore at $t=0$ the solution is $(x_0,y_0,z_0)$.

\begin{figure}
\centering
\subfloat{\includegraphics[scale=0.2]{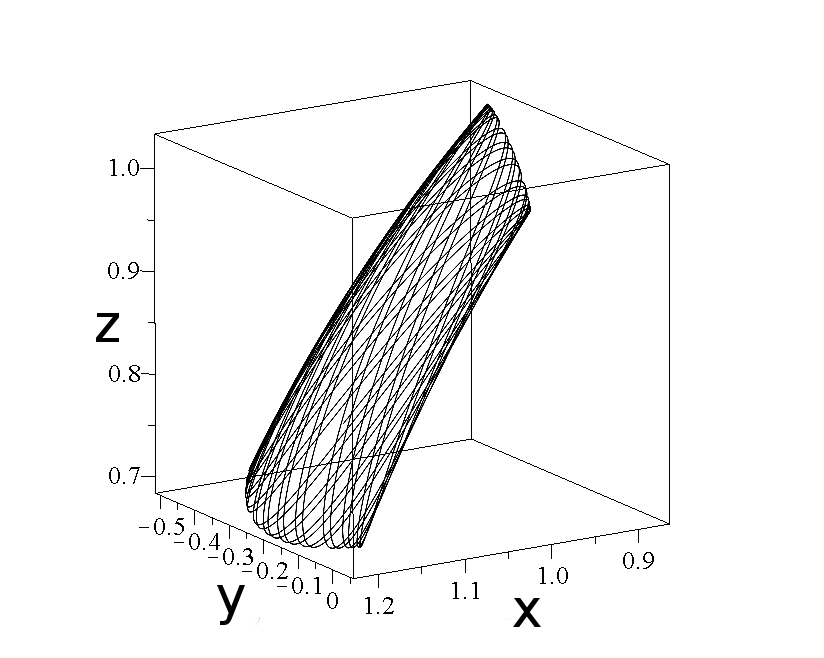}}[a]\label{original}
\subfloat{\includegraphics[scale=0.2]{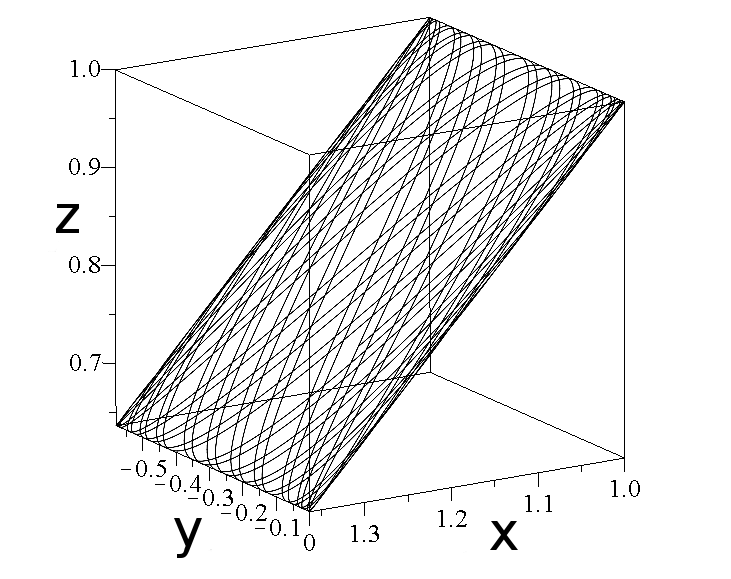}}[b]\\
\subfloat{\includegraphics[scale=0.16]{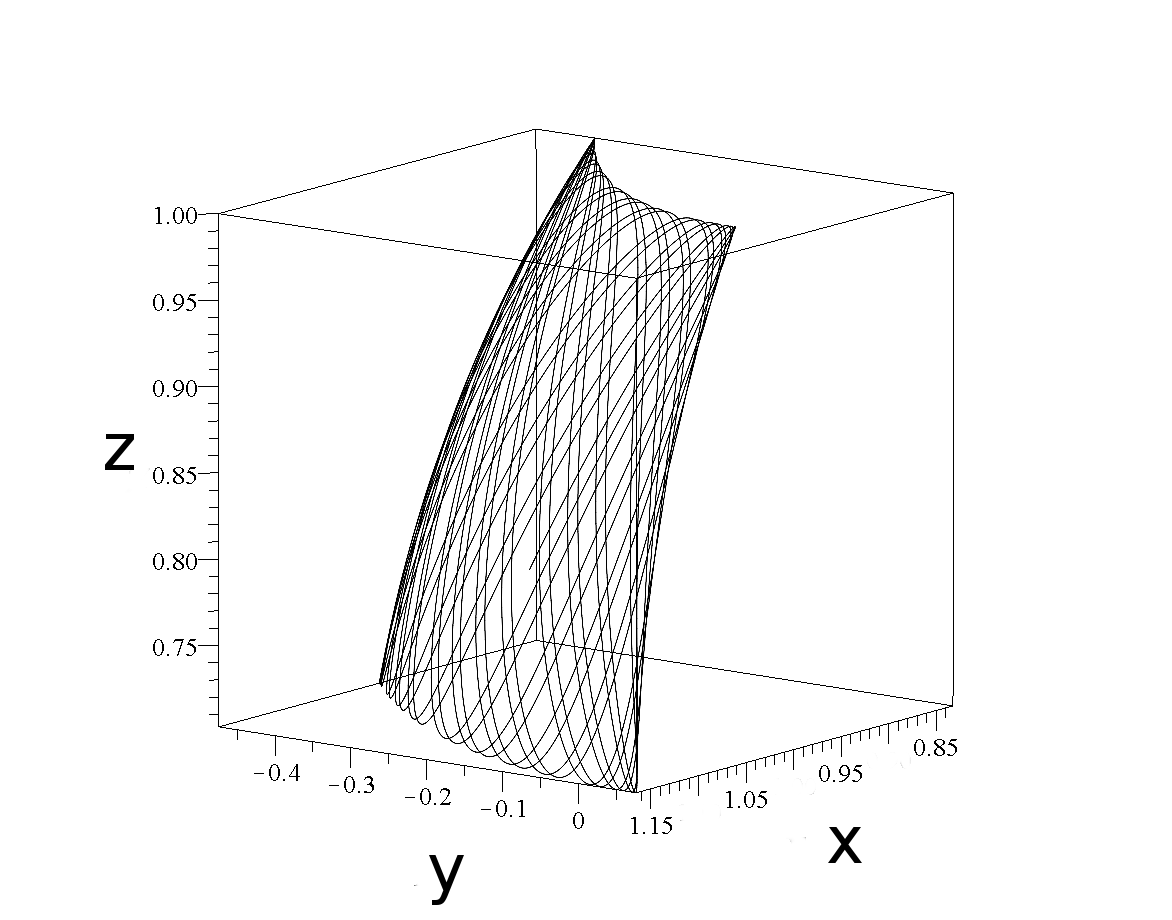}}[c]
\caption{(a)An ordered trajectory on the spherical surface with $R=\sqrt{2}$, computed numerically by the exact Bohmian system, with $b=c=0.1$ (error $\sim 10^{-5}$) (b) the first order solution (c) The second order solution  approaches better the real solution (initial condition $(x(0)=1, y(0)=0, z(0)=1, t\in[0,200]$).}
\label{ordereds}
\end{figure}

An example of an ordered trajectory on the sphere is given in Fig.~\ref{ordereds}. The first figure shows the numerical solution of the Bohmian equations \ref{bohmeqs1}-\ref{bohmeqs3}. By running backwards in time the integration, we find the initial conditions with an error $\sim 10^{-5}$. The second figure shows the first order solution provided by Eqs. \ref{first}-\ref{third}  which in general deviates a little from the spherical surface. In fact from these equations we derive $x_0(x-x_0)+y_0(y-y_0)+z_0(z-z_0)=0$ and this represents a plane.
Furthermore we find $x^2+y^2+z^2=x_0^2+y_0^2+z_0^2$ therefore the solutions satisfy the  integral (\ref{sfairaol}). Thus in this case we have an exact integral (\ref{sfairaol}) for any values of $b$ and $c$ and two more formal integrals for small $b$ and $c$.
We have obtained second order approximations (Fig.~(8c)) of the solution that give an even better approach to the numerical solution of Fig.~(8a). Approximations of higher order can be computed by using a computer-algebraic program, in the same way as in the 2-d case developed in \cite{contopoulos2008ordered}. However if the factors b and c are large most orbits become chaotic and the approximate solutions are no more valid. On the other hand on the spherical surface (\ref{sfairaol}) we find also chaotic orbits.

\begin{figure}[ht]
\centering
\includegraphics[scale=0.5]{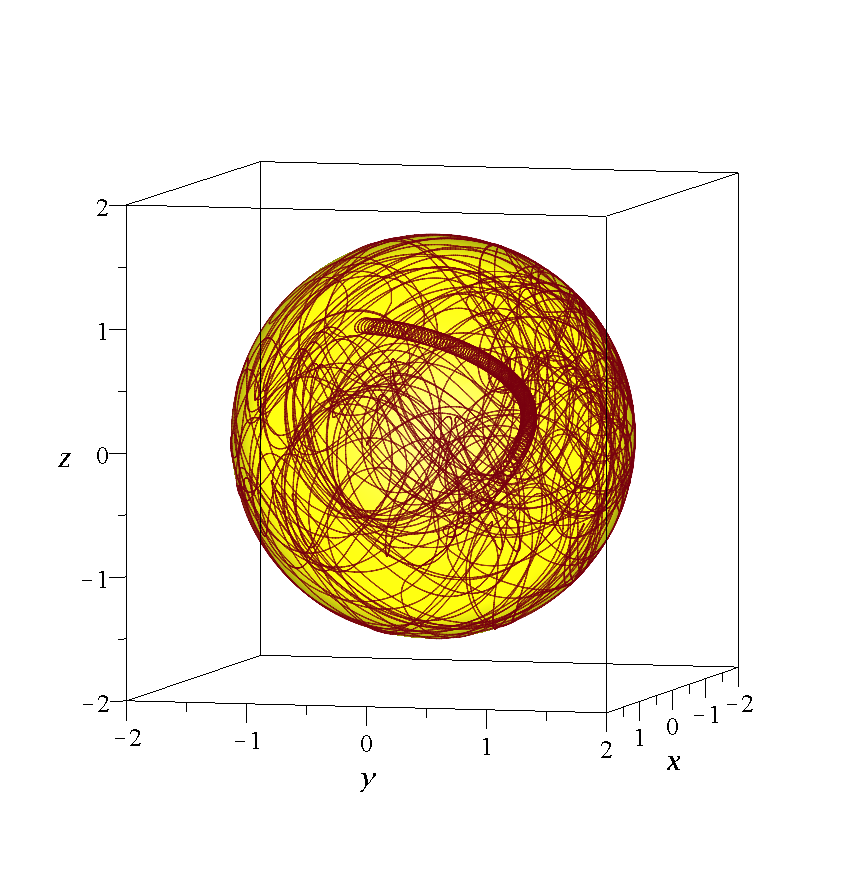}
\caption{A chaotic trajectory starting close to the nodal point of $t=4$ on the sphere with radius $R=1.6274$ for the interval $t\in[4,1000]$. We observe the Bohmian vortices around it for the first $4.5$ time units. After that time, the nodal point accelerates abruptly and the trajectory gets derailed, following a completely irregular path on the spherical surface. Here we have one exact integral and no formal integrals. ($x(4)=1.297366, y(4)=-.262989,z(4)=0.946631$)}
\label{sphch}
\end{figure}

A chaotic orbit starting close to a nodal point-Xpoint complex (in the case with $a=b=c=1/\sqrt{3}$) is shown in Fig.~\ref{sphch}. This orbit starts close to a nodal point and forms, for a certain time interval, loops around the nodal point. As the nodal point proceeds downwards to the right the loops of the orbit also follow the moving nodal point. However after a certain time the nodal point is accelerated abruptly and then the particle cannot follow it and deviates considerably from the neigbourhood of the moving nodal point. From then as the orbit extends all over the surface of the sphere in a chaotic way.

Two ordered orbits are already shown in Fig.~\ref{np2}. These orbits enter into the regions covered by the nodal trajectory. However when the moving point is in these regions the nodal point is away. Thus the moving point does not approach the nodal point-X-point complex. On the other hand the chaotic orbits of Fig.~\ref{np2} approach the nodal point-X-point complex. We notice also that the left chaotic orbit overlaps the region covered by the ordered orbit. However the two orbits do not overlap in time.
\subsubsection{Pear-shaped surfaces}
\begin{figure}
\centering
\includegraphics[scale=0.45]{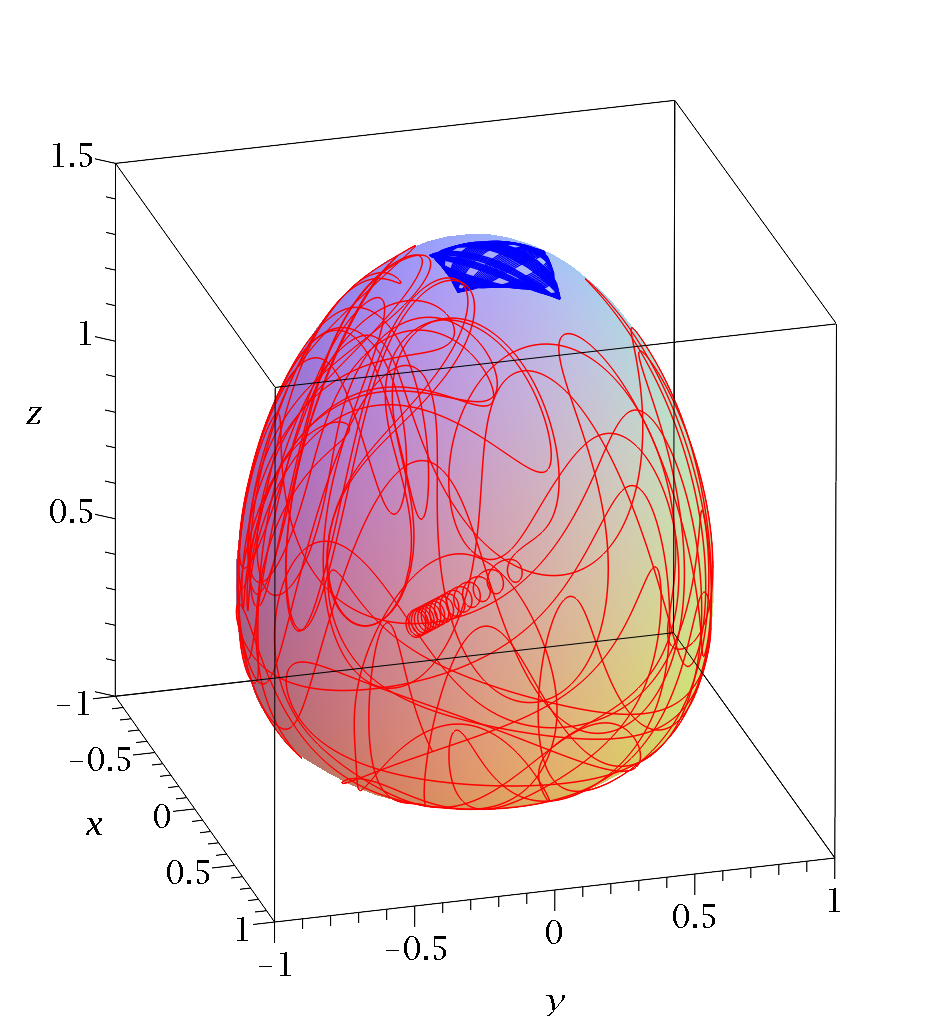}
\caption{Ordered  and chaotic  Bohmian trajectories on the pear shaped surface with $C=1$. \Big[($x(1)=0.209550, y(1)=0.11, z(1)=1.45)$ and $(x(1)=0.691268, y(1)=-.408571, z(1)=0.723687)$ respectively. The ordered trajectory has  error $\sim 10^{-6}$.\Big].}
\label{axladi_troxies}
\end{figure}

\begin{figure}
\centering
\subfloat{\includegraphics[scale=0.55]{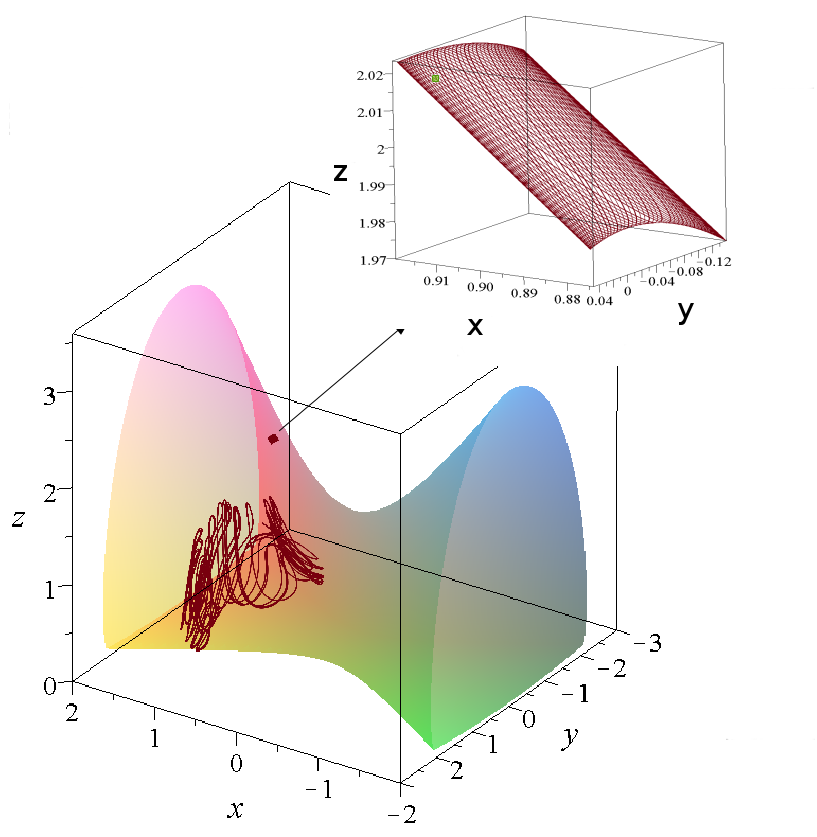}}\label{anoikti_troxia_f}
\caption{Ordered and chaotic Bohmian trajectories on the open surface ($x(1) = .8943744433, y(1) = 0, z(1)=2$ and $x(1) = 0, y(1) = 1/\sqrt{2}, z(1)=1$ correspondingly and $t\in[1,100]$). The ordered orbit is shown also in magnified form and its error is of order $10^{-6}$.}
\label{orderedsopen}
\end{figure}

Similar results apply also to the other partially integrable cases.
In Fig. \ref{axladi_troxies} we present an ordered and a chaotic orbit on the same surface in the pear-shaped case. In  particular the orbits starting close to a nodal point-Xpoint complex are chaotic (Compare the chaotic orbits of Figs. \ref{sphch} and \ref{axladi_troxies}). On the other hand orbits that remain far from the nodal point are ordered. (E.g. the orbit on the top of the pear of Fig.~\ref{axladi_troxies} remains in the empty region that is not covered by the orbit of the nodal point in Fig.~\ref{np4x}). Similarly in the case of the open integral surface we have both ordered and chaotic orbits as seen in Fig. \ref{orderedsopen}.

\clearpage

\subsection{Non Integrable Cases}
In the absence of partial integrability Bohmian trajectories are free to wander around the 3-d space and the study of their evolution becomes significantly more difficult.
As an example, we study the case
\begin{align}
\Psi(\vec{x},t)=a\Psi_{0,0,0}(\vec{x},t)+b\Psi_{1,0,1}(\vec{x},t)+c\Psi_{0,1,2}(\vec{x},t).
\end{align}

\begin{figure}[h]
\centering
\subfloat{\includegraphics[scale=0.21]{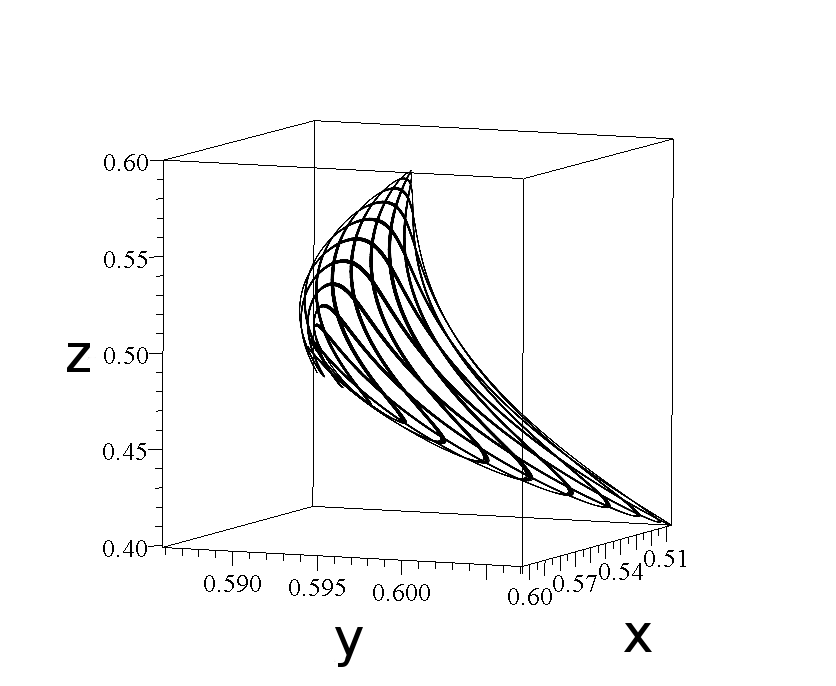}}[a]\label{original}
\subfloat{\includegraphics[scale=0.19]{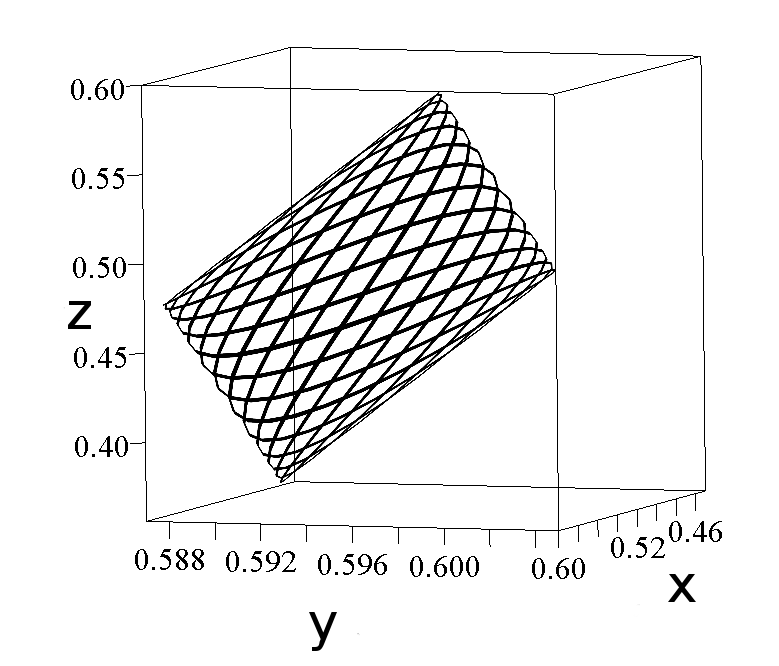}}[b]\\
\subfloat{\includegraphics[scale=0.2]{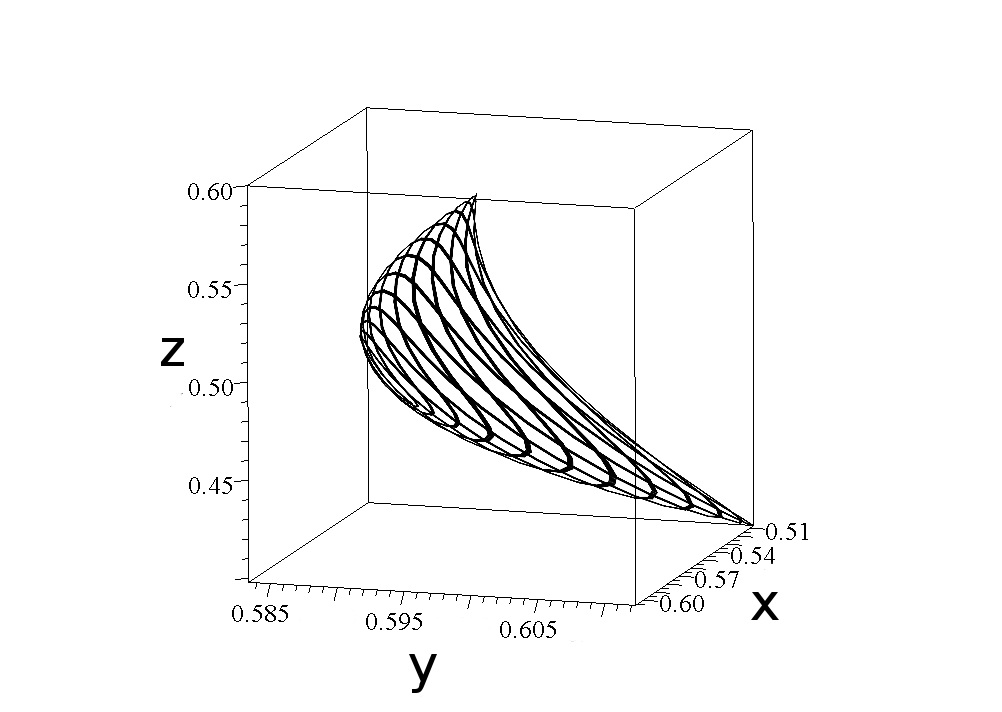}}[c]
\caption{(a)An ordered trajectory in the non-integrable case, computed numerically by the exact Bohmian system, with $b=c=0.1$ (error $\sim 10^{-5}$). (b) The first order solution which gives a plane. (c) The second order solution which approximates the curvature of the trajectory $(x(0)=0.6, y(0)=0.6, z(0)=0.6$, $t\in [0,100]$).}
\label{ordered_full_chaos_fail}
\end{figure}

\begin{figure}
\centering
\subfloat{\includegraphics[scale=0.37]{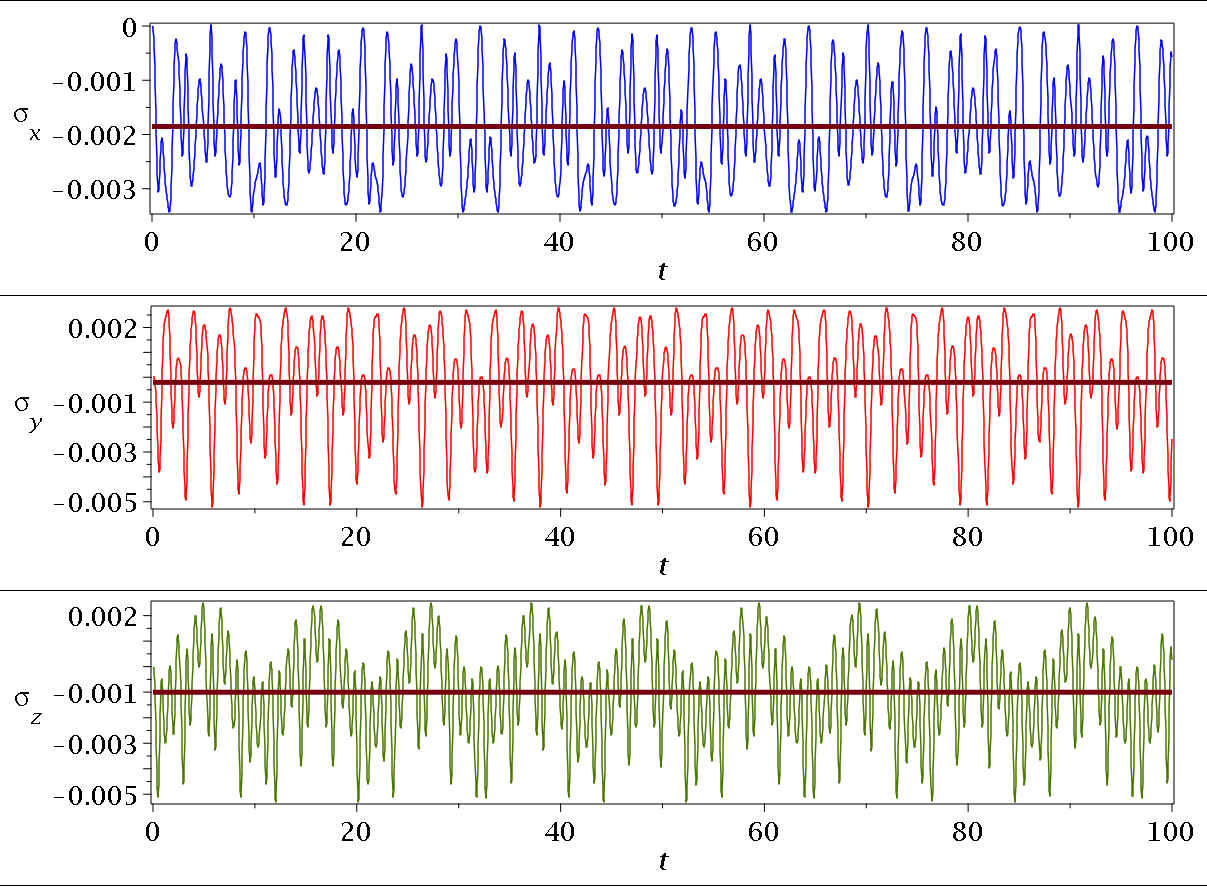}}
\caption{The deviations $\sigma_x, \sigma_y, \sigma_z$ of the coordinates $x, y, z$ calculated with the second order expansion from those calculated numerically. We observe that the average error is $\sim 10^{-3}$. With higher order perturbation theory one expects to find more accurate solutions.}
\label{errors}
\end{figure}

\begin{figure}[ht]
\centering
\subfloat{\includegraphics[scale=0.55]{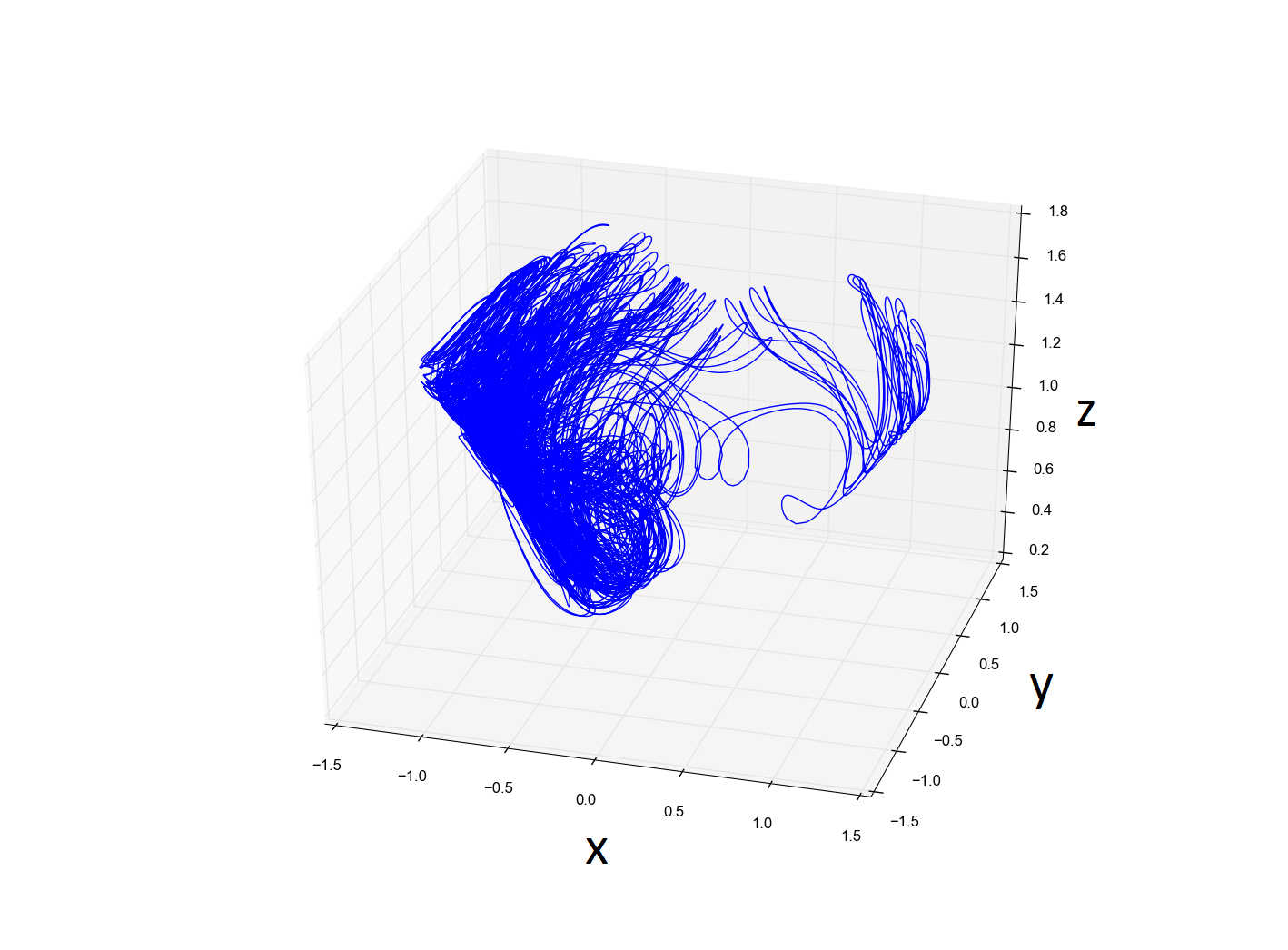}}
\caption{A non-integrable Bohmian trajectory for $t\in[0, 1000]$ with $x(0)=-0.500, y(0)=0.330, z(0)=0.732$. In this case there is no formal or exact integral of motion ($a=b=c=1/\sqrt{3}$).}
\label{fullchaos}
\end{figure}

If we treat $b,c$ as small perturbations, we are able to apply again regular perturbation theory as in the case of the sphere. The solutions of the first order approximation are:
\begin{align}
&x=x_0+2\,{\frac {\sqrt {\omega_1\omega_3}z_{0}\, \left( \cos \left(
 \left( \omega_1+\omega_3 \right) t \right) -1 \right) }{a \left(
\omega_1+\omega_3 \right) }}b\label{firstc}
\\&
y=y_0+2\,{\frac {\sqrt {\omega_2} \left( {z_{0}}^{2}\omega_3-1/2 \right)
 \left( \cos \left(( \omega_2+2\,\omega_3)t \right)  -1
 \right) }{ \left( \omega_2+2\,\omega_3 \right) a}}c\label{secondc}\\&
 z=z_0+2\,{\frac {\sqrt {\omega_1\omega_3}x_{0}\, \left( \cos \left(
 \left( \omega_1+\omega_3 \right) t \right) -1 \right) }{a \left( \omega_
1+\omega_3 \right) }}b+4\,{\frac { \left( \cos \left(  \left( \omega_2+2\,\omega_3 \right) t
 \right) -1 \right) z_{0}\,\omega_3\,\sqrt {\omega_2}y_{0}}{ \left(
\omega_2+2\,\omega_3 \right) a}}c\label{thirdc}
\end{align}

The quantum numbers of this wavefunction do not satisfy the necessary conditions for the existence of partial integrability. We note that the first order solutions (\ref{firstc}-\ref{thirdc}) have trigonometric terms with only two frequencies $\omega_1+\omega_2$ and $\omega_2+2\omega_3$. The same is true also for the higher order terms. If we solve Eqs. (\ref{firstc}-\ref{secondc}) we find $\cos{(\omega_1+\omega_3})t$ and $\cos{(\omega_2+2\omega_3})t$ as functions of $x$ and $y$ respectively. Introducing these values into Eq. \ref{thirdc} we find the equation of a plane
\begin{align}
(z_0^2\omega_3-1/2)[z_0(z-z_0)-x_0(x-x_0)]-2y_0(y-y_0)z_0^2\omega_3=0
\end{align}
which is time-independent. Then we can find $\cos{(\omega_1+\omega_3})t$ and $\cos{(\omega_2+2\omega_3})t$ in higher approximations and introducing them into the equation for $z$ we find a time-independent formal integral $z=z(x,y)$. Any truncation of this formal integral represents approximately a surface on which takes place the motion for small values $b$ and $c$. Therefore in the general non-integrable case we find three formal integrals of motion, two of them time-dependent and one time-independent.

An example is shown in Fig.~\ref{ordered_full_chaos_fail}. The numerical solution is given in Fig.~12a and the first and second order approximatons are given in Figs. 12b and 12c. The first order approximation is on a plane but the second order approximation approaches well the form of the exact orbit. Indicatively, in Fig.~\ref{errors} we present the deviations between the numerical solution and the second order approximation, as well as their average value, which is of order $10^{-3}$. 

However, these integrals are only formal and there are small deviations both on the approximate surface and away from that surface. These deviations are much larger when the values of $b$ and $c$ become larger. An example is shown in Fig.~\ref{fullchaos}, where $a=b=c=1/\sqrt{3}$. This orbit is chaotic and is not restricted on any surface.

\section{Conclusions}
In this paper we studied various aspects of Bohmian trajectories in 3 dimensional systems, working with a system corresponding to 3-d harmonic oscillators.

\begin{enumerate}
\item{Namely we considered wavefunctions consisting of the sum of 3 eigenfunctions multiplied by constands $a,b,c$ with $|a|^2+|b|^2+|c|^2=1$ of the form (\ref{form}). The system with $
b=c=0$ is completely integrable, i.e. it has three exact solutions (integrals) $x=x_0, y=y_0, z=z_0$. For small values of $b$ and $c$ we find always two independent formal integrals of motion which represent ordered orbits. This result is similar to a corresponding result in the 2-d case \cite{contopoulos2008ordered} where we had found one formal integral of motion.}
\item{However, when $b$ and $c$ are large these formal integrals are no more valid and the orbits become chaotic. In fact chaos is introduced when an orbit approaches the nodal point-X-point complex \cite{Efth2009,Tzemos2016}. Orbits starting close to a nodal point make spirals around it for some time but then they deviate considerably from it. This result is also similar to that of the 2-d case.}
\item{Many systems are partially integrable, i.e. they have one exact integral of motion. In such cases the orbits lie on the integral surfaces that are either closed (e.g. spherical or pear shaped) or open. The nodal points move again on such surfaces and their form leaves empty regions on which lie most of the ordered orbits. In partially integrable systems, besides the exact integral which is valid for all values of the amplitudes $a, b$ and $c$, we have one more formal integral if $b$ and $c$ are sufficiently small.}
\item{In non-integrable systems orbits there are again three formal integrals of motion for small amplitudes $b$ and $c$  that define approximately a surface. The ordered orbits lie again close to such a surface (e.g. in the first approximation the orbits evolve on a plane). However in general the orbits are chaotic and they cover a  3-d region of space.}
\item{We applied regular perturbation theory in order to find approximate integrals of motion for ordered trajectories in the spherical case and in the non-integrable case.
The existence of integrability in Bohmian Mechanics is of both mathematical and physical interest:}
\item{Partial integrability simplifies significantly the computations of the basic structures of the Bohmian flow. In particular in the case of the spherical and pear-shaped surfaces, we introduced new coordinates based on their symmetry in order to study the shape of the nodal point-X-point complex and the forms of the orbits (see also the Appendix).}
\item{Partial integrability implies the coexistence of order (studied by formal integrals) and chaos on the same integral surface, something that affects the rate of quantum relaxation and needs to be further studied.}
\end{enumerate}

\section*{Appendix: The trajectory of the nodal point}
The next step after the succesfull finding of a nodal point for a certain time $t_0$ is to trace its motion in space, something that might be a demanding task conceptually and computationally, because one needs to make a `1-1' map between the 3-d nodal point of a nodal line at time $t$ and a 3-d nodal point of a nodal line at time $t+dt$, something difficult especially in the cases where multiple nodal points coexist at a given time or when the nodal point tends to infinity. However we present 3 ways to proceed:
\begin{enumerate}
\item{The first method is based on the fact that close to a given nodal point, the Bohmian trajectories lie on a plane orthogonal to the locally tangent vector of the nodal line. This we call `F-plane' \cite{falsaperla2003motion}. The calculation of an F-plane is described in the algorithm below.
\begin{enumerate}
\item{We identify the nodal curve for a given time $t$. In general the nodal line is parameterized by one of the spatial coordinates of the nodal point, say $z_{nod}$. Namely it is
$\vec{r}_N(z_{nod})=f_1(z_{nod})\hat{i}+f_2(z_{nod})\hat{j}+z_{nod}\hat{k}.$
The tangent vector to the nodal curve at a certain nodal point with $z_{nod}=c$ is
\begin{align}
\vec{t}=\Bigg(\frac{\partial f_1(z_{nod})}{\partial z_{nod}}\hat{i}+
\frac{\partial f_2(z_{nod})}{\partial z_{nod}}\hat{j}+\hat{k}\Bigg)_{z_{nod}=c}
\end{align}
}
\item{We find the F-plane, that is perpendicular to the tangent vector:
$(\vec{r}-\vec{r}_N)\cdot \vec{t}=0$
Consequently for $z_{nod}=c$
\begin{align}\label{f_eq}
\Big(x_1-f_1(c)\Big)f_1'(c)+\Big(x_2-f_2(c)\Big)f_2'(c)+\Big(x_3-c\Big)=0
\end{align}
where $f_i'=\frac{\partial f_i(z_{nod})}{\partial z_{nod}}$.
Equation~(\ref{f_eq}) describes the F-plane for a given time $t$ and $z_{nod}=c$. The velocity of the nodal point lies on the F-plane.
We expect now the nodal point at time $t'=t+dt$ to remain very close to the F-plane of time $t$ if $dt$ is very small. Then we can find its coordinates $x_{nod}',y_{nod}',z_{nod}'$ and apply again the above algorithm in order to calculate the F-plane at time $t'$. The successive application of this procedure will trace the motion of a given nodal point in space. This method is applicable whether partial integrability exists or not. Actually it is the only method available in the absence of partial integrability.}\end{enumerate}}
\item{The second method is applicable to partially integrable cases. By knowing the integral surface upon which lie the trajectories and consequently the nodal points, we are able to give the integral surface with respect to one spatial coordinate (say $z_{nod}$) and time, $g(z_{nod},t)=C$.
Then we can solve this equation numerically with a root finding method like Newton-Rapshon
by taking as an initial guess of $t+dt$ the nodal point of time $t$. The efficiency of this method relies heavily on the accuracy of the guess for the position of the nodal point at time $t+dt$. It can fail if the nodal point has large velocity.}
\item{The third method differs from the second one only in that it uses the differential form of the integral surface. Consequently one needs again to solve an initial value problem for $z_{nod}$. According to our experience this procedure is quite faster than the previous one.}
\end{enumerate}

\textbf{Acknowledgements}: This work was supported by the Research Committee of the Academy of Athens. It has been conducted in the frame of the project of RCAAM "Study of the dynamical evolution of the entanglement and coherence in
quantum systems".
The authors want to thank Dr. C. Efthymiopoulos for many useful comments.

\bibliographystyle{unsrt}       
\bibliography{bibliography.bib}   

\end{document}